\documentclass[showpacs,preprintnumbers,amsmath,amssymb,aps,prb]{revtex4-1}
\usepackage{mathrsfs}

\usepackage{graphicx}
\usepackage{times}
\usepackage{dcolumn}
\usepackage{bm}
\usepackage{array}
\usepackage{hyperref}
\usepackage{amsmath}
\usepackage{rotating}
\usepackage{graphics,color}
\usepackage{xcolor,pict2e,curve2e}

\usepackage[normalem]{ulem}
\usepackage{todonotes}
\usepackage{soul}

\newlength{\figwidth} 

\newcommand{\beq}{\begin{equation}}

\newcommand{\beql}[1]{\begin{equation}\label{#1}}

\newcommand{\eeq}{\end{equation}}
\newcommand{\bsp}{\begin{aligned}}
\newcommand{\esp}{\end{aligned}}

\newcommand{\Eq}[1]{Eq.~(\ref{#1})}

\newcommand{\Table}[1]{Table.~(\ref{#1})}

\newcommand{\Fig}[1]{Fig.~\ref{#1}}
\newcommand{\Figure}[1]{Figure~\ref{#1}}

\begin{document}

\title{General point dipole  theory for periodic metasurfaces: magnetoelectric  scattering lattices coupled to planar photonic  structures}

\author{Yuntian Chen}
\affiliation{School of Optical and Electronic Information, Huazhong University of Science and Technology, Wuhan, 430074, China}

\author{A. Femius Koenderink}
\affiliation{Center for Nanophotonics, FOM Institute AMOLF, Science Park 104, 1098 XG Amsterdam, The Netherlands}

\date{\today}
\begin{abstract}
We study semi-analytically the light emission and absorption properties of arbitrary stratified photonic structures with embedded two-dimensional magnetoelectric point scattering lattices, as used in recent plasmon-enhanced LEDs and solar cells. By employing dyadic Green's function for the layered structure in combination with Ewald lattice summation to deal with the particle lattice, we develop an efficient method to study the coupling between planar 2D  scattering lattices of plasmonic, or metamaterial point particles, coupled to layered  structures. Using the `array scanning method' we deal with localized sources. Firstly, we apply our method to light emission enhancement of dipole emitters  in slab waveguides, mediated by plasmonic lattices. We benchmark the array scanning method against a reciprocity-based approach to find that the calculated radiative rate enhancement in k-space below the light cone shows excellent agreement. Secondly, we apply our method to study absorption-enhancement in thin-film solar cells mediated by periodic Ag nanoparticle arrays. Lastly, we study the emission distribution in k-space of a coupled waveguide-lattice system. In particular, we explore the dark mode excitation on the plasmonic lattice using the so-called Array Scanning Method.  Our method could be useful for simulating a broad range of complex nanophotonic structures, i.e., metasurfaces, plasmon-enhanced light emitting systems and photovoltaics.
\end{abstract}


\maketitle

\section{Introduction}

In nanophotonics,  periodically structured media and periodic texturing of devices are of large interest both for fundamentally new approaches to controlling light, and for applications in spectroscopy \cite{Willets2007ARPC}, solid-state lighting \cite{Schnitzer,Vuckovic,Okamoto,Yeh,Henson,BarnesJLT1999,RigneaultOL1999,Lozano2013}, and photovoltaics \cite{LareAPL2012,Ferry2010OE}. Current   developments in fundamental electromagnetics research particularly focus on plasmonic lattices \cite{Kravets,Auguie2008PRL,LareAPL2012,Ferry2010OE,Lozano2013,Schokker2014,TormaRPP2015}, metamaterials \cite{Gansel2009NatPhys,SersicPRL2009}, and metasurfaces\cite{YuScience2011,YuNatureM2014,Kivshar2015LPR}. Plasmonic lattices are two-dimensional subwavelength or diffractive periodic structures of metal nanoparticles, such as spheres, cylinders, bars or pyramids, in which the large scattering strength of plasmonic localized resonances is combined with collective scattering effects to give strong local fields\cite{AbajoOE2006,Auguie2008PRL}. Currently these are of large interest due to the possibility of strong coupling between localized modes and diffraction conditions \cite{TormaRPP2015}, and the possibility of strong coupling of the resulting hybrid photonic modes with embedded emitters. In metamaterials the individual inclusions are generalized from the plasmonic case to also support magnetic, or bi-anisotropic resonances, as is the case in split rings\cite{SersicPRB2011,Jingbia1}, and helices\cite{Gansel2009NatPhys}.  Metasurfaces, finally, are 2D optically thin polarizable particle arrays, in which individual magneto-electric inclusions may be rationally designed to be non-identical, and if periodic, often take the form of frequency  selective surfaces, where a single unit cell may contain many magneto-electric inclusions \cite{Radi2015prapplied}. For gradient structuring of the meta-atoms in a unit cell, such type of metasurfaces can be useful in the wavefront engineering for creating ultrathin optical components \cite{YuScience2011,YuNatureM2014} such as  lenses, wave plates and holograms. Metasurface have further been studied extensively as perfect absorbers \cite{Landy2008PRL}, for asymmetric transmission \cite{Fedotov06,Menzel10} , and interfacing   bound modes and free propagation photons \cite{Schnitzer,Vuckovic,Okamoto,Yeh,Henson,BarnesJLT1999,RigneaultOL1999}.

\begin{figure}
\centerline{\includegraphics[width=0.8\textwidth]{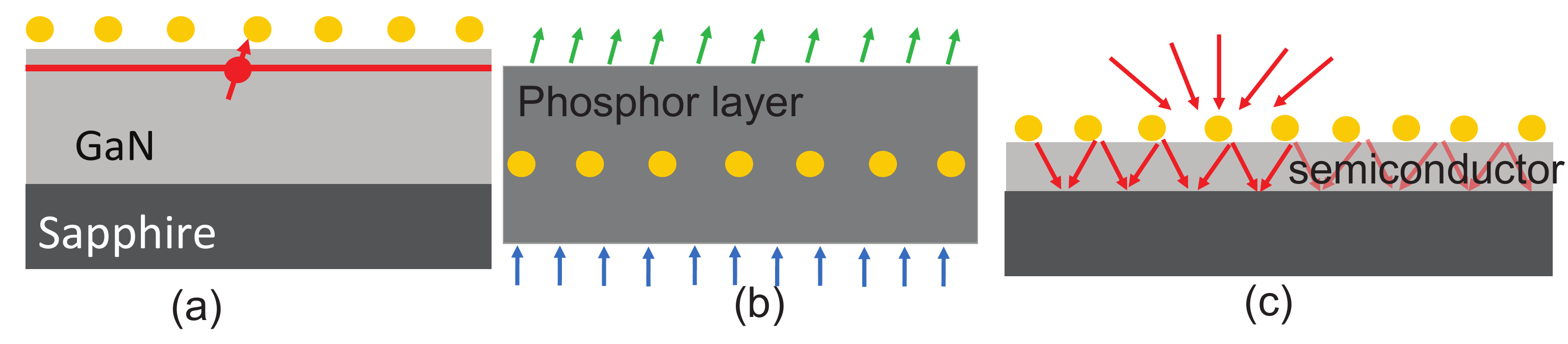}}
\caption{\label{fig1} Sketches of canonical scenarios that our theory can address. (a) Plasmonic lattice patterned on a GaN material based LED structure to improve outcoupling; (b) Remote phosphor for white light emission from LED systems, pumped by a blue LED. Plasmonic texturing can improve absorption, and redirect emission.  (c) thin film solar cell with texturing to improve light incoupling.}
\end{figure}

Our paper is motivated by a host of applications of periodic plasmonic lattices and metasurfaces that have been demonstrated in recent experiments. In particular, periodic texturing of devices with metal structures has been proven to be highly successful in solid-state lighting applications \cite{Henson,BarnesJLT1999,RigneaultOL1999,DavidRPP2012,Lozano2013,IidaAIP2015},  and for improved photovoltaics \cite{LareAPL2012,Ferry2010OE}.  Figure~\ref{fig1} summarizes several of the main scenarios.  In solid state lighting,  emission is typically generated in a high index GaN or In$_x$Ga$_{1-x}$N semiconductor light emitting device (LED) that is grown on a sapphire substrate. Light that can not be extracted due to total internal reflection and guided modes of the high index layer, can be collected if the LED is textured with a dielectric or plasmonic pattern (Fig.~\ref{fig1}a).   A second scenario appears in current white light LEDs, where emission is generated by using an efficient blue LED as pump for a phosphorous wavelength conversion layer \cite{Lozano2013}. Here the challenge is to minimize the use of phosphorous material by using plasmonic texturing as a means to enhance blue light absorption, and to optimize (directional )phosphor emission extraction (Fig.~\ref{fig1}b). A third application scenario is in silicon photovoltaics. Crystalline and amorphous silicon solar cells suffer from long absorption lengths, and high interface reflection. Periodic structuring by plasmon particles can be used to reduce reflections, and to cause diffractive coupling to guided modes in the solar cell \cite{CatchpoleAPL2008,CatchpoleOE2008,LareAPL2012,Ferry2010OE}. Thereby absorption  can be enhanced, and cell thicknesses can be reduced (Fig.~\ref{fig1}c).   In all these scenarios, the periodic plasmon structure is placed in, or on,  a nontrivial stratified dielectric system with a complex mode structure.

While the  relevance of periodic plasmonic lattices and metasurfaces embedded in stratified systems is large,  no simple design or analysis tool is available. Results usually are compared to brute force numerical  calculations such as FDTD or FEM modeling \cite{Fedotov015,Pors2013OE,Chen2010PRB,Chen2010OE}. This method provides accurate full-wave simulation results, but involves extensive computational efforts.  Since for a \emph{given} structure, cataloguing the angle- and frequency dependent response usually takes of order 100 seconds per angle and frequency, the challenge of testing many designs is herculean. Moreover, the mechanism of the modes hybridization, or  the interplay between the particle resonances, lattice resonances, and guided stack modes is usually very difficult to retrieve from full wave numerical calculations.
It is  desirable to have a physically transparent and conceptually simple model that can be applied to plasmonic and metamaterial lattices embedded in arbitrary stratified systems. In this work we present a dipole lattice model that can deal with   arbitrary stratified systems. Considerable efforts have been devoted to developing dipole lattice models \cite{Radi2015prapplied,SersicPRB2011,PerPRB2013} for plasmonic and magneto-electric scatterer systems, where retardation as well as the cross coupling between electric dipole and magnetic dipole moments are taken into account through Ewald lattice summation.  We recently extended this to lattices in front of a simple interface \cite{KwadrinPRB2013,KwadrinPRB2014}.  In this paper, we treat dipolar lattices arbitrarily positioned in arbitrarily stratified structures, as sketched in the possible scenarios in \Fig{fig1}. Our model is not limited to any specific range of periodicities or scattering strengths of the lattice, and can thus handle metamaterials, diffractive gratings, and any regime in between.  Moreover, using the "array scanning method" we show how to deal with single point emitters driving the periodic systems.

In  section II, we give a comprehensive description of the model, which includes the construction of the dyadic Green function of magnetoelectric point scatter lattices coupled to planar stratified stacks, the array scanning method, the local density of optical states, and far field emissions patterns. In section III,  we present results for solid-state lighting and photovoltaic scenarios. Section IV presents the conclusions of the paper.

\section{Theory}
Our overall strategy to build a theoretical model is the following.  First, we set up a general theory for interaction between dipolar scatterers embedded in a complex dielectric environment. The scatterers are captured in a polarizability, while the interactions are quantified by a Green function.  We focus on stratified systems,  meaning that the Green function can be built in an angular spectrum representation, separating TE and TM waves. As next step, we determine how to solve for interactions in infinite lattices using Ewald summation, on the proviso that the excitation field is not localized, but rahter has some definite parallel momentum, as would be provided in a plane-wave scattering experiment. Finally we discuss how to obtain the response to localized driving from superposition of the momentum-space solution by the Array Scanning Method (ASM). Section~\ref{subsec:G} introduces the Green function, section~\ref{subsec:pw} sets up the infinite-lattice dipole problem in momentum space, and section~\ref{subsec:dipole} introduces the array scanning method. In section~\ref{subsec:farfield} we discuss an asymptotic expression for evaluating far field radiation, and for calculating the  local density of states via reciprocity.
\subsection{Dyadic Green's function for a single magnetoelectric point dipole\label{subsec:G}}
For self-consistency, we start from the definition of the Green's tensor \cite{LeePiers2007,Novotny2006}, which reads,
\begin{subequations} \label{MaxwellGG}
\begin{align}
    [\nabla  \times  \nabla  \times  - k_0^2 \bar{\bm {\varepsilon}}_r (\bm r)\bar{\bm{\mu}}_r(\bm r) ] \bar {\bm{G}}^{ee} (\bm r,\bm r' )= \bar {\bm {I}}_0\delta(\bm r-\bm r'), \\
    [\nabla  \times \nabla  \times  - k_0^2 \bar{\bm {\varepsilon}}_r (\bm r)\bar{\bm{\mu}}_r(\bm r) ] \bar {\bm{G}}^{he} (\bm r,\bm r' )= \bar {\bm {I}}_0 \nabla  \times \delta(\bm r-\bm r'), \\
[\nabla  \times  \nabla  \times  - k_0^2 \bar{\bm{\varepsilon}}_r(\bm r)\mu_r (\bm r)] \bar {\bm{G}}^{hh} (\bm r,\bm r' )= \bar {\bm {I}}_0\delta(\bm r-\bm r'), \\
[\nabla  \times  \nabla  \times  - k_0^2 \bar{\bm{\varepsilon}}_r(\bm r)\mu_r (\bm r)]  \bar {\bm{G}}^{eh} (\bm r,\bm r' )=- \bar {\bm {I}}_0 \nabla  \times \delta(\bm r-\bm r'),
\end{align}
\end{subequations}
where $\bar {\bm{G}}^{(e/h) e}$ denotes the electric/magnetic field Green's tensor from an electric point source, and $\bar {\bm{G}}^{(e/h) h}$ denotes electric/magnetic field Green's tensor from a magnetic point source. The fields provided by an object with electric current $ \bm j_{e}(\bm r)$ and magnetic current  $\bm j_{h}(\bm r)$ are given by
\begin{eqnarray} \label{free_green}
\begin{aligned}
\bm E (\bm r) & =i \omega_0 \mu_0   \int  \bar { \bm{ G } }^{ee}(\bm r, \bm r') \bar{\bm{\mu}}_r(\bm r')\bm j_{e}(\bm r')d\bm r' \\ &+\frac{1}{\bar{\bm{\epsilon}}_{r} (\bm r)}\int \bar { \bm{ G } }^{eh}(\bm r, \bm r')  \bar{\bm{\epsilon}}_{r} (\bm r') \bm j_{h}(\bm r')d\bm r' \\
 \bm H (\bm r) &=i \omega_0 \varepsilon_0  \int  \bar { \bm{ G } }^{hh}(\bm r, \bm r') \bar{\bm{\varepsilon}}_r(\bm r') \bm j_{h}(\bm r')d\bm r' \\ &+ \frac{1}{\bar{\bm{\mu}}_{r} (\bm r)} \int \bar { \bm{ G } }^{he}(\bm r, \bm r')  \bar{\bm{\mu}}_{r} (\bm r') \bm j_{e}(\bm r')d\bm r'.
\end{aligned}
\end{eqnarray}
In this work we focus on scatterers that can be modelled as point dipoles, requiring that they are moderate in size (order $\lambda/2\pi$) and sufficiently spaced (spacing at least the particle radius).  Although the examples that we present all deal with plasmonic scatterers that only have an electric polarizability,  for completeness we also incorporate a magnetic dipole response in the model. Thereby, the model also applies to many metasurfaces. For a point scatterer, the induced electric and magnetic dipole moments  $\bm P$ and $\bm M$ can be expressed in terms of induced currents through $\bm j_{e}(\bm r)=-i \omega_0 \bm P \delta(\bm r- \bm r _0)$, $\bm j_{h}(\bm r)=-i \omega_0 \bm M \delta(\bm r- \bm r _0)$. Using normalized fields $[\bm e, \bm h]=[\bm E, Z_0\bm H]$ and normalized dipole moments  $[\bm p, \bm m]=[\frac{\bm P}{4\pi \epsilon_0}, \frac{Z_0 \bm M}{4\pi \mu_0}]$ (tantamount to using CGS units), one can simplify the fields due to an arbitrary electric and magnetic dipole scatterer as
\beq\label{pm_eh2013}
\begin{pmatrix}
   \bm e \\
   \bm h \\
 \end{pmatrix}  =   4 \pi k_0\left( \begin{array}{cc}
                 k_0  \bar{\bm\mu}_r^{e}  \bar { \bm{ G } }^{ee}   &    -i  \frac{\bar{\bm{\epsilon}}(\bm r_0)}{\bar{\bm{\epsilon}}(\bm r)}  \bar { \bm{ G } }^{eh} \\
                  -i \frac{\bar{\bm{\mu}}(\bm r_0)}{\bar{\bm{\mu}}(\bm r)}  \bar {\bm{ G }}^{he}     &    k_0\bar{\bm\varepsilon}_r^{e}   \bar { \bm{ G } }^{hh}
                \end{array}\right) \begin{pmatrix}
\bm p \\
   \bm m \\
 \end{pmatrix}.
\eeq
In this work we focus on lattices embedded in planar photonic structures, consisting of an arbitrary multilayer stack. In this case, to construct the multilayer Green function one  uses plane waves, decomposed into TM and TE polarization  for each given propagation direction, as specified by parallel momentum $\mathbf{k}_{||}$ (momentum component in the plane parallel to the interfaces) as a complete basis to describe any field in space.
By convention,  this set of plane waves is expressed through a set of basis functions $\bm M(\bm k_{||})$ defined as describing the electric field of TE waves, and a dual set ($\bm N(\bm k_{||})$) that describes the electric field of TM waves.  One set of basis functions can be derived from the other  since  $ \bar { \bm{ G } }^{ee} = \bar { \bm{ G } }^{hh}$,  the $\bm M(\bm k_{||})$ functions not only describe the E-field of TE waves, but also describe the TM-wave $H$-fields. From the relation between E-field and H-field in source-free regions, $  \bar { \bm{ G } }^{eh}(\bm r, \bm r')  = -\nabla  \times  \bar { \bm{ G } }^{hh}(\bm r, \bm r') $,  one finds $ \nabla  \times \bm M(\bm k_{||}) = k_0 \bm N(\bm k_{||})$. Conversely, one can also derive $  \nabla  \times \bm N(\bm k_{||}) = k_0 \bm M(\bm k_{||})$, where curl  operates on the coordinates of detection point $\bm r$.
This relation facilitates a straightforward computation of all sub-tensors of $ \bar { \bm{ G } }$ in terms of $\bm M(\bm k_{||})$  and $\bm N(\bm k_{||})$.
The appendices report the functions  $\bm M(\bm k_{||})$  and $\bm N(\bm k_{||})$ for free space (Appendix A, essentially plane waves),  and for arbitrary stratified systems. In any multilayer structure, the functions  $\bm M(\bm k_{||})$ and $\bm N(\bm k_{||})$  can be found by applying the transfer matrix formalism, that introduces the wave-vector dependent fresnel coefficients into the problem. Appendix C generalizes the aforementioned reported recipe \cite{KlimovPRA2005, HartmanJCP,Cheng1986} for Green's tensor $  \bar { \bm{ G } }^{ee}$ to  the complete description of the $\bm E$ and $\bm H$ fields induced by a single magneto-electric point scatter.

\subsection{2D  lattice and plane wave excitation\label{subsec:pw}}
We now turn to the case of a 2D periodic magneto-electric point scatter lattice coupled to a slab waveguide  under a plane wave illumination. We consider the response of a 2D lattice (lattice vectors $\bm R_{\bm l}=m\bm a_{1}+n\bm a_{2}$) of point dipoles of polarizability $\bar {\bm \alpha}$ to a driving  field of parallel wave vector $\bm k_{||}$, where we index lattice sites with the label $\bm l =[m, n]$.   The response is set by  solving  a self-consistent coupled equation given by
\beq\label{polarizability}
 \begin{pmatrix}
   \bm {p}_{\bm l}\\
   \bm {m}_{\bm l}
 \end{pmatrix}  =\bar {\bm \alpha}   \left[  \left( \begin{array}{c}
               \bm e^{\bm k_{||}}(\bm R_{\bm l})  \\
             \bm h^{\bm k_{||}}(\bm R_{\bm l})
                \end{array}\right) +  \sum \limits_{ \bm l' \neq \bm l} \bar{\bm {g}}(\bm R_{\bm l},\bm R_{\bm l'})
 \begin{pmatrix}
   \bm p_{\bm l'} \\
   \bm m_{\bm l'}
 \end{pmatrix}  \right].
\eeq
where $\bar {\bm \alpha}$ is the dynamic polarizability tensor,  $\bar{\bm {g}}(\bm R_{\bm l},\bm R_{\bm l'})=4 \pi \left( \begin{array}{cc}
                 k_0^2 { \bm{ G } }^{ee} (\bm R_{\bm l},\bm R_{\bm l'})& -ik_0\bar { \bm{ G } }^{eh} (\bm R_{\bm l},\bm R_{\bm l'})\\
                  -ik_0\bar { \bm{ G } }^{he} (\bm R_{\bm l},\bm R_{\bm l'}) & k_0^2 \bar { \bm{ G } }^{hh}(\bm R_{\bm l},\bm R_{\bm l'})
                \end{array}\right)
$, and where  $\bm p_{\bm l}$ ($\bm m_{\bm l}$ ) represents the  induced electric (magnetic) dipole moment in the point scatter seated at $\bm R_{\bm l}$. By using the stratified medium dyadic Green's function  $\bar{\bm g}(\bm r,\bm r')$,  given in Appendix \Eq{slab_green}, we ensure that all multiple scattering interactions between the induced dipole moments $\bm{p}_{\bm l}$ that are mediated by both propagating and guided modes  are included.    The driving field $ \bm e^{\bm k_{||}}(\bm R_{\bm l})$ ($ \bm h^{\bm k_{||}}(\bm R_{\bm l})$) is understood to be the field solution  for the full stratified system in absence of the nanoscatterers in response to an incident plane wave $\bm E_{in}^0$ of parallel momentum $\bm k_{||}$, as specified in \cite{Urbach1998,Jung2011}.  According to  de Vries et al. \cite{Vries1998}, the dynamical polarizability tensor  $\bar{\bm {\alpha}}$ of a point scatter can be written as $
\bar{\bm{\alpha}} =  \left[\bar{\alpha}_{static}^{-1}-4\pi k^2 \bar { \bm{ G } }_b (\bm r _0,\bm r _0) \right]^{-1}$, where  $\bar{\alpha}_{static}$ is a quasistatic polarizability (no radiation damping), and where $\bar { \bm{ G } }_b $ is the background Green's tensor at the position of  the point scatter itself. A regularization procedure needs to be carried out to remove the singularity in $\real\bar { \bm{ G } }_b $. Since this regularization is not   specific to the stratified system, we refer to Ref.~\onlinecite{Vries1998} for its derivation. For a point scatterer in free space, the routine boils down to taking solely the imaginary part of $\bar { \bm{ G } }_b $, which accounts for the radiation damping to fulfill the optical theorem. In a  complex system, one corrects the  polarizability tensor of a point scatter by adding  to the inverse static polarizability $4\pi k^2i \mathrm{Im} \bar { \bm{ G } }_{free}$ (referring to the Green function of homogeneous space filled with the medium directly surrounding the scattering), plus the scattered part $4\pi k^2{\bar { \bm{ G } }_s(\bm r_0, \bm r_0)}$ of the Green's tensor that contains all the multiple reflections in the multilayer stack, and is regular. Mathematically, one adds to the inverse of  static polarizability, i.e., $\alpha^{-1}=\alpha_{static}^{-1}- \frac{2ik^3}{3}\bar {\bm {I}}_0-{\bar { \bm{ G } }_s(\bm r_0, \bm r_0)}$.

Returning to \Eq{polarizability}, the translation invariance of the coupled lattice-waveguide system implies that a  Bloch wave form  $\left[ \bm{p}_{\bm l}, \bm{m}_{\bm l} \right]^T=e^{i \bm k_{||} \cdot \bm R_{\bm l}} \left[ \bm{p}_{\bm 0}, \bm{m}_{\bm 0} \right]^T$ can be used to simplify the infinite set of coupled equations to a single equation
\begin{equation}
 \begin{pmatrix}
   \bm p_{\bm 0} \\
   \bm m_{\bm 0}
 \end{pmatrix} =[\bar {\bm \alpha}^{-1}- \bm g^{\neq}]^{-1}  \begin{pmatrix}
   \bm E_{in} (\bm R_{\bm 0})  \\
   \bm H_{in} (\bm R_{\bm 0})
 \end{pmatrix},
\end{equation}
where $\bar {\bm g}^{\neq}$ is the lattice summation of the stratified system Green's function without self-term, given by $\bar{\bm g}^{\neq}=\sum \limits_{\bm l\neq \bm 0}\bar{\bm g}(\bm R_{\bm 0},\bm R_{\bm l}) e^{i \bm k_{||} \cdot \bm R_{\bm l}}$, $\bm R_0$ is the coordinates of the particle labeled by $\bm l=[0, 0]$ on the lattice. Even for dipole interactions in free space, it is a   challenge  to calculate the lattice sum  $\bar{\bm g}^{\neq}$, which  can be tackled using Ewald summation techniques. We have adapted the recipe  of Linton~\cite{Linton2010} for summations involving just scalar Green functions in free space to deal also with dyadic Green functions of arbitrary stratified systems.  To this end, we reformulate the total Green's tensor of stratified structures as $\bar{\bm g}(\bm r,\bm r')=\bar{\bm g}_0(\bm r,\bm r')+\bar{\bm g}_s(\bm r,\bm r')$, where $\bar{\bm g}_0$ and $\bar{\bm g}_s$ are the free space term and the scattered term due to the interfaces, respectively.  The explicit form of $\bar{\bm g}_0$ and $\bar{\bm g}_s$   can be seen in Appendix(A-C).  This separation of  $\bar{\bm g}(\bm r,\bm r')$ leads to
\beq\label{deomT0000}
\begin{aligned}
\bar{\bm{\alpha}}^{-1}-\bar{\bm g}^{\neq}(\bm k_{||})  & =  \bar {\bm {\alpha}}_{static}^{-1}-\frac{2ik^3}{3} - \sum\limits_{\bm l } \bar { \bm{ g } }_s(\bm R_{\bm 0},\bm R_{\bm l}) e^{i \bm k_{||} \cdot \bm R_{\bm l}}\\ &-   \{ \sum\limits_{\bm l}  \bar { \bm{ g } }_0(\bm R_{\bm 0},\bm R_{\bm l})  e^{i \bm k_{||} \cdot \bm R_{\bm l}}-\bar { \bm{ g } }_0(\bm R_{\bm 0},\bm R_{\bm 0})\}.
\end{aligned}
\eeq
The `free' term (entire term in $\{.\}$) in \Eq{deomT0000} can be efficiently calculated using  Ewald summation. We use the recipe reported in Ref~\onlinecite{PerPRB2013,KwadrinPRB2014}. At first sight, the summation over $\bar { \bm{ g } }_s$ in the third term in \Eq{deomT0000} must be much more daunting, since the scattered part of the Green function is formulated as a Sommerfeld integral over parallel wavector containing the $\mathbf{M}$ and $\mathbf{N}$ modes of the multilayer. Remarkably, the sum of poorly convergent complex integrals   simplifies to a rapidly converging sum of just the integrand at select $\bm k _{||}$ that rapidly converges  due to the evanescent nature of the higher diffraction orders. Taking one block in $\bar { \bm{ g } }_s$ for an interface as an example,  we explicitly formulate  $ \sum\limits_{\bm l} \bar { \bm{ G } }_s^{ee}(\bm R_{\bm 0},\bm R_{\bm l})e^{i \bm k_{||} \cdot \bm R_{\bm l}} $ for the example of a lattice placed on top of a slab waveguide as follows,
\beq\label{slabgreen22s0000}
\begin{aligned}
& \sum\limits_{\bm l } \bar { \bm{ G } }_s^{ee}(\bm R_{\bm 0},\bm R_{\bm l}) e^{i \bm k_{||} \cdot \bm R_{\bm l}}=  \sum\limits_{\bm l } \iint_\infty^\infty dq_xdq_y \frac{i}{8 \pi^2 k_{z1}} \\ & [{\bm M_{+}\otimes \gamma_{s}^{1,23} \bm M'_{+} } + { \bm N_{+}\otimes \gamma_{p}^{1,23}\bm N'_{+}}]   e^{-i\bm q_{||} \cdot (\bm R_{\bm l}-\bm R_{\bm 0})}e^{i \bm k_{||} \cdot \bm R_{\bm l}}.
\end{aligned}
\eeq
Here we refer to the Appendix for a listing of the quantities involved - essentially $\gamma_{s,p}^{1,23}$ accounts for the polarization-dependent fresnel coefficients of the multilayer system, generalized to extend to large $k_{||}$. The crucial realisation is that in this complicated integrand there is no dependence on $\mathbf{R}_l$ except in the very last term, i.e., the phase factor $e^{i (\bm k_{||}-\bm q_{||}) \cdot \bm R_{\bm l}}$.  Hence one can pull the summation over lattice sites into the integrand and apply the completeness relation  $\sum\limits_{\bm l} e^{i (\bm k_{||}-\bm q_{||}) \cdot \bm R_{\bm l}}=\frac{(2\pi)^2}{A}\sum\limits_{\bm l }  \delta((\bm k_{||}-\bm q_{||})-\bm{\mathscr{G}}_{\bm l} )$ (where $\bm{\mathscr{G}}_{\bm l}= m \bm{\mathscr{G}}_{x}  +n\bm{\mathscr{G}}_{y}  $, $\bm{\mathscr{G}}_{x}  $ and $\bm{\mathscr{G}}_{y}  $ are reciprocal lattice vectors and $A$ denotes the real-space unit cell area)  to convert the real space summation of integrals in \Eq{slabgreen22s0000}  into a reciprocal space sum with no complex  integrals as follows,
\beq\label{sumslabgreen22s0000}
\begin{aligned}
& \sum\limits_{\bm l } \bar { \bm{ G } }_s^{ee}(\bm R_{\bm 0},\bm R_{\bm l}) e^{i \bm k_{||} \cdot \bm R_{\bm l}}= \frac{(2\pi)^2}{A}\sum\limits_{\bm l} \frac{i}{8 \pi^2 k_{z2}} \\ &[{\bm M_{+}\otimes \gamma_{s}^{1,23} \bm M'_{+} } + { \bm N_{+}\otimes \gamma_{p}^{1,23}\bm N'_{+}}]e^{i\bm k_{||} \cdot \bm R_{\bm 0}}|_{(\bm q_{||}=\bm k_{||}+\bm{\mathscr{G}}_{\bm l})}.
\end{aligned}
\eeq
In  \Eq{sumslabgreen22s0000}, the sum  can be easily truncated (despite the existence of guided modes with a very long in-plane real-space range), due to the fact that each term carries a  $e^{i2k_z d}$ ($d$ denotes the distance of particle center to its nearest neighboring  interface), which exponentially decreases as $|{\bm k}_{||}+{\mathscr{G}}_{\bm l}|$ is large enough. The same rationale holds for all blocks in $\bar { \bm{ g } }_s$.
Once the induced moment $\bm{p}_{\bm 0}$ and $\bm{m}_{\bm 0}$ is obtained, it is straightforward to calculate the total  field at any location  as
\begin{equation}\label{pwfields}
\begin{pmatrix}
    \bm E_{total}^{\bm k_{||}}(\bm r)  \\
   \bm H_{total}^{\bm k_{||}}(\bm r)
 \end{pmatrix} = \begin{pmatrix}
    \bm E_{slab}^{\bm k_{||}}(\bm r)  \\
   \bm H_{slab}^{\bm k_{||}}(\bm r)
 \end{pmatrix}+ \sum \limits_{\bm l}\bar{\bm g} (\bm r,\bm R_{\bm l}) e^{i \bm k_{||} \cdot \bm R_{\bm l}}  \begin{pmatrix}
   \bm p_{\bm 0} \\
   \bm m_{\bm 0}
 \end{pmatrix}.
\end{equation}

\subsection{Point dipole excitation\label{subsec:dipole}}
We proceed to discuss the coupled  lattice-waveguide under a point source excitation by employing the array scanning method (ASM) reviewed, for instance,  by Capolino et al. in \cite{Capolino2007}. The use of Bloch's theorem, and the Ewald lattice summation technique, as we presented sofar, restricts one to finding the response to extended driving field of definite parallel momentum, as it requires periodicity of the \emph{entire} system (geometry plus driving) under study. In principle, one can decompose  the driving by a single point source in plane waves, and thereby obtain the response of a lattice to local driving by superposition. More practical for implementation is the Array Scanning Method which states that  a single point source can be constructed from integration over phased arrays of sources~\cite{Capolino2007}.  Provided one can solve for any well-defined phased array $J_{\bm k_{||}}=\bm J\sum \delta(\bm r-(\bm r'+\bm R_{\bm l}))e^{i\bm k_{||}\cdot \bm R_{\bm l}}$ of definite parallel momentum, and the same periodicity as the lattice, one can retrieve the single point dipole source by integrating   $J_{\bm k_{||}}$ over the Brillouin zone, i.e., $\bm J(\bm r')=\frac{A}{(2\pi)^2}\int\limits_{BZ}  \bm J_{\bm k_{||}}d\bm k_{||}$. For each individual $\bm k_{||}$, the problem is directly compatible with the formulation that was set up for  plane wave excitation. When summing the results over all $\bm k_{||}$, one has to ensure each $\bm k_{||}$ receives   its own weighting, as given by the background Green's tensor. The central result of this approach is that the field due  to a  point source is
\beq\label{master_equation}
\bar { \bm{ G } } (\bm r,\bm r')  = \bar { \bm{ g } }_b (\bm r,\bm r') + \frac{A}{(2\pi)^2}\int\limits_{BZ} d \bm k_{||} \bar { \bm{ I } }(\bm k_{||},\bm r,\bm r_0)
\eeq
where $ \bar { \bm{ g } }_b (\bm r,\bm r')$ is the background  dyadic green function (i.e., that of the stratified system in absence of the lattice), the integrand of the second term $\bar { \bm{ I } } (\bm k_{||},\bm r,\bm r_0)$ is given by $\bar { \bm{ I } }(\bm k_{||},\bm r,\bm r_0)= \bm {g}^{=}(\bm r,\bm r_0)\bar{\bm Q}(\bm r_0,\bm r_0) \bm {g}^{=} (\bm r_0,\bm r')$, and $\bm g^{=}(\bm r_0, \bm r')=\sum \limits_{\bm l}\bar{\bm g}(\bm r_0, \bm r') e^{i \bm k_{||} \cdot \bm R_{\bm l}}$ is again a lattice-summed Green function, though now \emph{including} the self-term, i.e., $\bm l= \bm 0$ term. The interpretation of $\bar { \bm{ I } }(\bm k_{||},\bm r,\bm r_0)$ is that the first term gives direct radiation from the source without interaction with the lattice. The second term contains the emission scattered via the lattice, and is the term containing the array scanning. The integrand of second term has three factors, that should be read from right to left:
\begin{enumerate}\item $\bm g^{=}(\bm r_0, \bm r') =\sum \limits_{\bm l}\bar{\bm g}(\bm r_0, \bm r') e^{i \bm k_{||} \cdot \bm R_{\bm l}} $,   gives the field strength at the particle lattice site $\bm r_0$ excited by the array of discrete current source $\bm J e^{i\bm k_{||} \cdot \bm R_{\bm l}}$ located at $\bm r'+\bm R_{\bm l}$,
\item $\bar{\bm Q}=[\bar{\bm{\alpha}}^{-1}-\bar{\bm g}^{\neq}(\bm k_{||})  ]^{-1}$ quantifies how much the lattice is polarized by the phased array source $\bm J e^{i\bm k_{||} \cdot \bm R_{\bm l}}$, \item $\bm g^{=}(\bm r, \bm r_0)=\sum \limits_{\bm l}\bar{\bm g}(\bm r, \bm r_0) e^{i \bm k_{||} \cdot \bm R_{\bm l}}$,  describes the field strength at $\bm r$ of field radiated by the polarized lattice.
\end{enumerate}

The calculation of  $\bar{\bm {g}}^{=}(\bm r,\bm r_0)$  and $\bar{\bm {g}}^{=} (\bm r_0,\bm r')$ can be performed by applying the completeness relation  as discussed for   $\bar{\bm {g}}^{\neq}(\bm r_0,\bm r_0)$. A major difference, however, is that  now  the source  point $\bm r_0$ might be in a \emph{different} layer than the lattice, as opposed to the lattice sum appearing for particle-particle interactions. An apparent problem is that the Green's tensor introduces   $1/k_{nz}$ terms, which introduce a singularity at the light line for medium $n$. Detailed analysis shows that the $1/k_{1z}$ in the Green's tensor cancels out with the   $k_{1z}$ proportionality of the fresnel transmission coefficient, which removes the numerical singularity.  As example, consider a three-layer system, with an  emitter in layer $1$ ($\bm r'=[0, 0, z'] $), while the particle lattice is in layer 2, so that  we need a  corresponding dyadic Green tensor that accounts for the field propagation  across the interface through a generalized fresnel transmission coefficient.  We take only one block in $\bar { \bm{ g } }_b (\bm r,\bm r')$, e.g., $ \bar { \bm{ G } }_{slab}^{{EE,21}}(\bm r_0, \bm r') $, to illustrate this useful cancellation result, which is given by
\beq\label{slabgreen21}
\begin{aligned}
&  \bar { \bm{ G } }_{slab}^{{EE,21}}(\bm r_0, \bm r')  = \iint dk_xdk_y \frac{i}{8 \pi^2 } \{ \\
 & \frac{1}{\Delta_s}\{[\bm M_{-}(\bm k_{||}) + \gamma_{s,23}(k_{||})\bm M_{+}(\bm k_{||})]   T_{s,12}e^{i\psi}\}\otimes [\bm M'_{+}(\bm k_{||})]  \\
             &    -\frac{1}{\Delta_p}\{[ \bm N_{-}(\bm k_{||}) +  \gamma_{p,23}(k_{||})\bm N_{+}(\bm k_{||}) ]T_{p,12}e^{i\psi} \}\otimes [ \bm N'_{+}(\bm k_{||})] \}
\end{aligned}
\eeq
where $T_{s,ij}=\frac{ 2 } {k_{zi} +k_{zj} }$, and $T_{p,ij}=\frac{ 2 \sqrt{ \epsilon_i \epsilon_j} } {k_{zi} \epsilon_j+k_{zj} \epsilon_i}$, $\psi=(k_{2z}-k_{1z})d$. Importantly, such cancellations carry over to complex layered structures, meaning that numerical calculation greatly benefits from judicious ordering of terms.

As regards to  performing the integrations over the Brillouin zone of the type above,  we note that even if the $1/k_{nz}$ singularities are eliminated, singularities still can occur due to poles in the transmission coefficients  at ${\bm k}_{||}$ that match  guided modes of the stratified system. We implemented a finite element scheme to efficiently carry out the numerical integration. The basic idea is to discretize the integration domain into triangles on which a simple quadrature rule is evaluated, and to adaptively refine the triangle density to obtain convergence also at guided mode contributions.

\subsection{Far field asymptotic form and radiative LDOS\label{subsec:farfield}}
So far we discussed how to solve for the response of scattering particle lattices to extended and local excitation, without going into the question how to calculate far field observables. Using far field asymptotic forms of \Eq{pwfields} enables one to calculate   transmission, reflection, diffraction efficiency, and absorption with ease for any incident parallel momentum. The asymptotic form of \Eq{pwfields}  can be obtained \cite{Novotny2006}  by,
\beq\label{FarNearMapping}
\bm E_{\infty}(s_x,s_y,s_z)= - 2 i \pi k_z \bm E(k_x,k_y;z=z_r)\frac{e^{i k r}}{r}
\eeq
where $\bm s =[s_x,s_y,s_z]=[\frac{x}{r},\frac{y}{r},\frac{z}{r}]=[\frac{k_x}{k_0},\frac{k_y}{k_0},\frac{k_z}{k_0}]$ parametrizes the far-field `viewing direction', and where $\bm E(k_x,k_y;z=z_r)$  is the angular spectrum of the near field at an arbitrarily chosen  detection plane $z=z_r$ in the half-space that extends to the far field detector. For plane wave incidence with in plane wave vector $\bm k_{||}$,  the only Fourier components of $\bm E(k_x,k_y;z=z_r)$ contributing to the far field are at $[k_x,k_y]=[\bm k_{||}+\bm{\mathscr{G}}_{\bm l}]$ ,where $\bm l$ labels the discrete diffraction orders.

In this work we are particularly concerned with mapping the fractional, i.e., parallel wave-vector resolved, radiative local density of states, i.e.,  the radiated flux per solid angle in the far field that is due to a localized source. In the ASM scheme, the asymptotic form of the far field  Green's function can be approximated  from \Eq{master_equation} by  taking the angular spectrum at a reference plane $z=z_r$,
\beq\label{master_equationfar}
\begin{aligned}
& \bar { \bm{ g } } ^{far} ([k_x,k_y,z=z_r],\bm r')  =\bar { \bm{ g } }_b ([k_x,k_y,z=z_r],\bm r')  \\ &  + \frac{A}{(2\pi)^2}  {\bm {g}}^{=}_{\bm l =\bm 0}([k_x,k_y,z=z_r],\bm r_0)\bar{\bm Q}(\bm r_0,\bm r_0) {\bm {g}}^{=} (\bm r_0,\bm r')
\end{aligned}
\eeq
for radiation at an output angle given by $[k_x/k_0, k_y/k_0]=cos\theta[cos\phi,sin\phi]$. Here the term
$\bar { \bm{ g } }_b ([k_x,k_y,z=z_r],\bm r') $ represents the far field emitted directly by the source (no interaction with the lattice).
The second term accounts for radiation that appears via interaction with the lattice. It contains  three sub-terms multiplicating with each other:  the first sub-term $ \bar{\bm {g}}^{=} (\bm r_0,\bm r')$ accounts for how strongly the source drives the lattice, the second sub-term  $\bar {\bm Q}$ plays the role of polarizability, and the final sub-term $\bar{\bm {g}}^{=}_{\bm l =0}$ quantifies how each particle in the lattice scatters out. Explicit expressions  of $ {\bm {g}}^{=} (\bm r_0,\bm r')$  and $\bar {\bm Q}(\bm r_0,\bm r_0)$ are given in \Eq{master_equation}, while the  term $ {\bm {g}}^{=}_{\bm l =0}([k_x,k_y,z=z_r],\bm r_0)$ only takes the zero diffraction order, i.e., $\bm l=\bm 0$. We can use  \Eq{master_equationfar} to calculate the radiative local density of states by taking the  sampling points of $\bm k_{||}$ within the light cone, without the need of unfolding $\bm k_{||}$ into corresponding diffraction orders  for large pitches (larger than half wavelength), compared to sampling points taken from Brillouin zone.

An independent method to obtain the radiative local density of states is to use the reciprocity theorem. Reciprocity implies that the local field enhancement inside the waveguide upon far field plane wave illumination is equivalent to calculating the plane wave strength in the far field due to a localized source \cite{Janssen_OE_2010}.  Strictly, one needs to identify the local electric field enhancement in the waveguide, upon far field plane wave illumination  \textit{weighted } by a factor of $\frac{1}{k_z}$  with the fractional radiative local density of states enhancement. Hence, one  can calculate the far field emission pattern of  a localized source coupled to the lattice-waveguide  simply by calculating the local field at the emitter position due to a plane wave scattered by the coupled lattice-waveguide, incident from the direction of interest.  The reciprocity technique can be easily extended to a bulk light source, i.e., emitters randomly distributed inside the waveguide in a certain region, by integrating $|E|^2$ over the active area \cite{RodriguezPRB2014} to efficiently obtain the far field emission pattern. However, the disadvantage of the reciprocity method is that it will not allow to determine the guided-mode contribution to LDOS, nor of absorption induced by the lattice. Any contribution to LDOS (imaginary part of the full system Green dyadic)  that does not appear as far field power, is not accessible to the reciprocity method. Thereby both the ASM and reciprocity are useful, and provide complementary information.

\begin{figure}
\includegraphics[width=\textwidth]{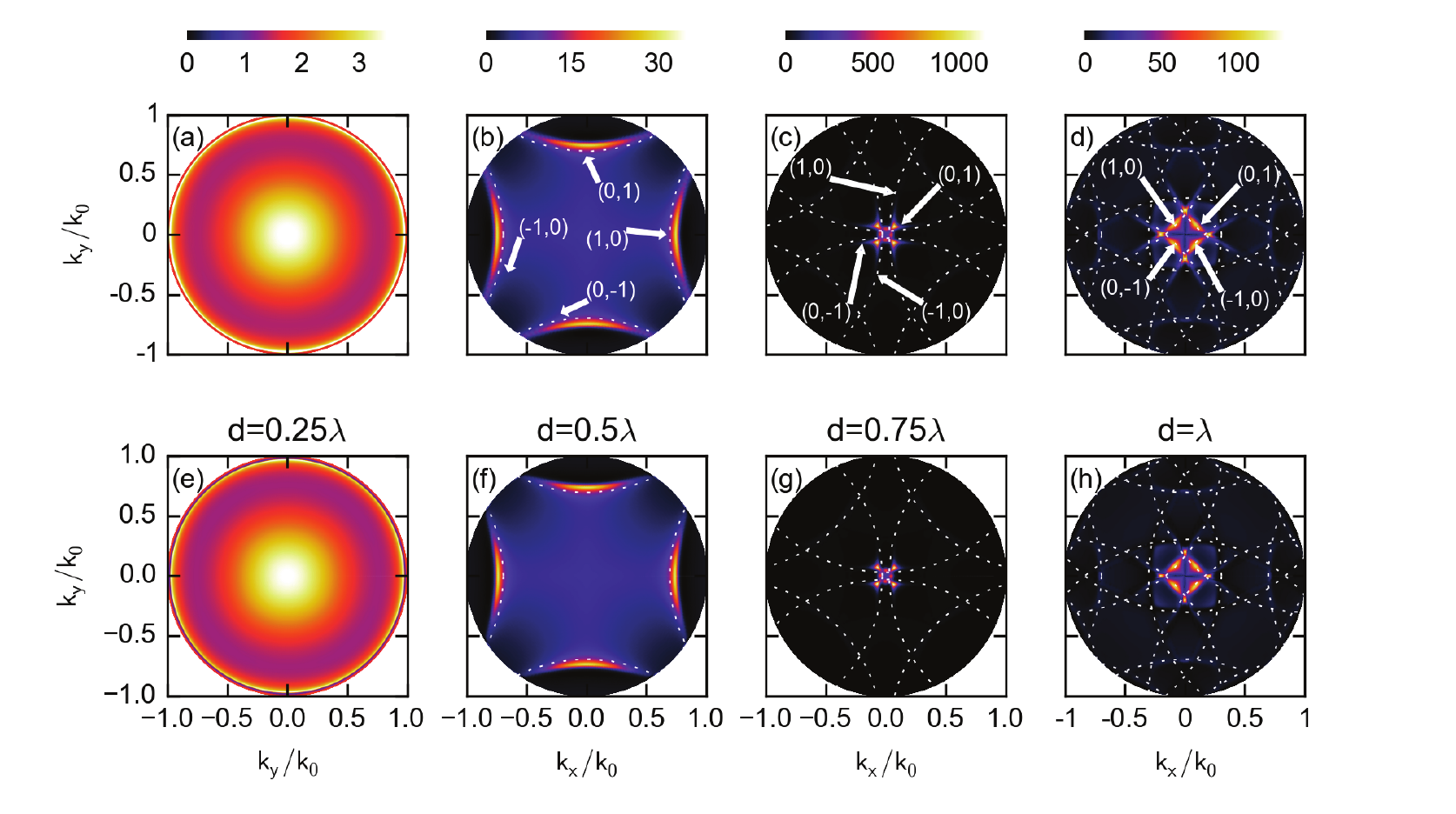}
\caption{\label{ASM_reciprocity} Angle-dependent far-field emission enhancement  of an electric dipole located  insidea  slab waveguide coupled to a plasmonic lattice, as sketched in \Fig{fig1} (a).  The host medium of the lattice is air, below which a waveguiding layer (index 1.6, thickness 0.2$\lambda_0$) is placed on a  perfect reflector. a-d (e-h) shows emission pattern calculated from  ASM method (reciprocity method) at four lattice constants, i.e., $1/4 \lambda_0$, $1/2 \lambda_0$, $3/4 \lambda_0$, $\lambda_0$ respectively. $\lambda_0$ is the vacuum wavelength. White lines are the folded slab waveguide inside the light cone.  In (b,c,d), the white number pairs denote the diffraction order.  The nanoparticle is taken as metallic sphere defined through an electrostatic polarizability of $(3.78+0.135i)\cdot 10^{-3}\lambda_0^3$ (typical for, e.g., a  80 nm aluminum particle in the green). The calculation is done for a single frequency determined  by $2\pi/\lambda_0$, and taking a source height $Z_0=1/10\lambda_0$.}
\end{figure}

%

\section{Results and discussion}
\subsection{Comparison of the far field pattern between reciprocity method and ASM scanning method}
We consider  the fractional  radiative  local density of optical states (FLDOS)  in a scenario similar to that recently explored by Lozano et al.~\cite{Lozano2013}and Rodriguez et al.~\cite{RodriguezPRB2014} for improvement of white light generation in thin phosphors that are excited by blue LEDs. In this scenario, a random ensemble of emitters in a thin layer of moderate index (about 1.5) is coupled to a lattice of plasmon particles to benefit from "diffractive "surface lattice resonances, and diffractive outcoupling of light emitted into waveguide modes. Experiments\cite{Lozano2013,RodriguezPRB2014} in particular showed highly directional emission (several degrees wide  emission cone) into angles matching grating outcoupling of waveguided modes. Here we present calculations of the fractional radiative LDOS for four different lattice pitches, to at the same time demonstrate the applicability of the point dipole model, and compare the ASM method to the reciprocity method. In \Fig{ASM_reciprocity} we consider the angle-dependent far-field emission enhancement  for a single emitter coupled to this waveguide-lattice structure, divided by the emission of a bare emitter  in the same slab structure without particles.

As stratified system we take a n=1.6 slab (the phosphor) positioned on a perfect electric conductor, and a thickness that is 1/5th of the free space emission wavelength (550 nm, green emission), i.e., 110 nm. As particles we take metal nanoparticles   placed on the slab, where the particles are defined through a static polarizability $\bar{\alpha}_{static}=(3.78+0.135i)
\cdot 10^{-3}\lambda_0^3$ (typical for, e.g., 80 nm Aluminum particles in the green). For this example we consider a single source position and orientation, although for actual phosphors one would need to perform incoherent averaging over all source positions and orientations.
For this example, we consider the source as an electric dipole moment perpendicular to the interfaces, so that only TM modes are excited. \Fig{ASM_reciprocity} shows both the angle-dependent far-field emission enhancement calculated according to the ASM method, with assistance of \Eq{master_equationfar}, as shown in \Fig{ASM_reciprocity} (a,b,c,d), and calculated according to reciprocity, i.e., as a near-field enhancement upon far-field TM-wave illumination, as shown in \Fig{ASM_reciprocity} (e,f,g,h). Results are shown for four different lattice constants, chosen as $ \lambda_0/4$ (subdiffractive), $\lambda_0/2$ (diffractive), $3\lambda_0/4$ (close to 2nd order Bragg diffraction for the waveguide mode) and $\lambda_0$.

In \Fig{ASM_reciprocity} (a,e), the lattice is effectively  a dense, non-diffractive metasurface,  as the pitch is chosen so small $(\lambda_0/4)$ that no diffraction contributes to the far field emission pattern.  For this lattice pitch,  the directional radiation enhancement  appears as a cone of about 30 degree opening angle. Here the reader should be warned that we plot enhancement, and   when dividing  to obtain photoluminescence \emph{enhancement}, one divides by the bare dipole emission, which is dark near the vertical for the chosen vertical dipole orientation.   As the lattice constant increases, the waveguide mode, with dispersion $\omega=|\mathbf{k}_{||}|n_{\mathrm{mode}}$ diffracts, and by diffraction is partially folded inside the light cone,    thus coupling out into the far field. For the larger pitch,  emission from the dipole that is emitted into the waveguide mode can hence be collected, and  is manifested as bright arcs with a radius of curvature given by the waveguide mode index in \Fig{ASM_reciprocity} (b,f), and the panel pairs (c,g) and (d,h). For larger pitch, the arcs approach, and show a anticrossing, as the condition of second order Bragg diffraction for the TM waveguide mode is passed. The bright arcs  in enhancement follow essentially the repeated-zone scheme folding of the waveguide slab mode dispersion by the plasmonic lattice (waveguide dispersion replicated every reciprocal lattice vector). Compared with the  pure waveguide folding as indicated by the white dash lines (unperturbed waveguide dispersion, plotted in repeated zone scheme dispersion, the bright arcs are broadened and slightly shifted. The broadening reflects a finite propagation length of the folded wave guide mode, due to the intrinsic Ohmic losses from the point scatters and the inclusion of radiation damping. As regards the shift of the mode to smaller mode index, we note that  the particles push the mode profile out of the high-index material thereby lowering the mode index, shifting features way from the drawn circles.  Qualitatively, i.e., in terms of showing sharp features that reflect the hybridized plasmonic-waveguide modes in a repeated zone scheme, the calculations are in good accord  with observations made by Lozano et al. and Schokker et al.~\cite{Lozano2013,Schokker2014}. In terms of magnitude of the enhancement, as shown in \Fig{ASM_reciprocity} (c,g), the predicted FLDOS enhancements are large, of order 100 (panel d) to 1000 (panel (c)). This large value is due to the fact that we use a  perpendicular dipole  for excitation, which in the bare layer case hardly radiates at small emission angles. Since our results are normalized to the bare dipole donut-shaped emission pattern,  this emphasizes emission at small $\mathbf{k}_{||}$ as a pronounced  enhancement.


A significant advantage of the reciprocity-based approach is that it requires only plane wave calculations, and that results can be obtained for all source positions and orientations in the unit cell in one calculation. This should be contrasted to the ASM method, that needs a full calculation per dipole position and orientation. However, the reciprocity based method can not separate out contributions to the local density of states that do not result in far field radiation, such as quenching and emission that remains trapped in waveguide modes. As we describe in section~\ref{subsec:dark}, the ASM does provide means to unravel these phenomena.
To highlight the calculation speed advantage of the ASM (the slowest of our two methods), over full wave simulations, we give a brief comparison of the computational time of our method and the full-wave simulation package Comsol. For ASM, each frame of the  calculation in  \Fig{ASM_reciprocity} (a-d) with  200$\times$100 computation points takes roughly 1 minute.  For a similarly complex structure, Comsol will take roughly 1 minute for a single computation point, on similar computer hardware. This significant difference in computational time signifies the efficiency of our method, as it allows to rapidly screen different geometries.

\subsection{Photovoltaic system}

\begin{figure}
\includegraphics[width=0.7\textwidth]{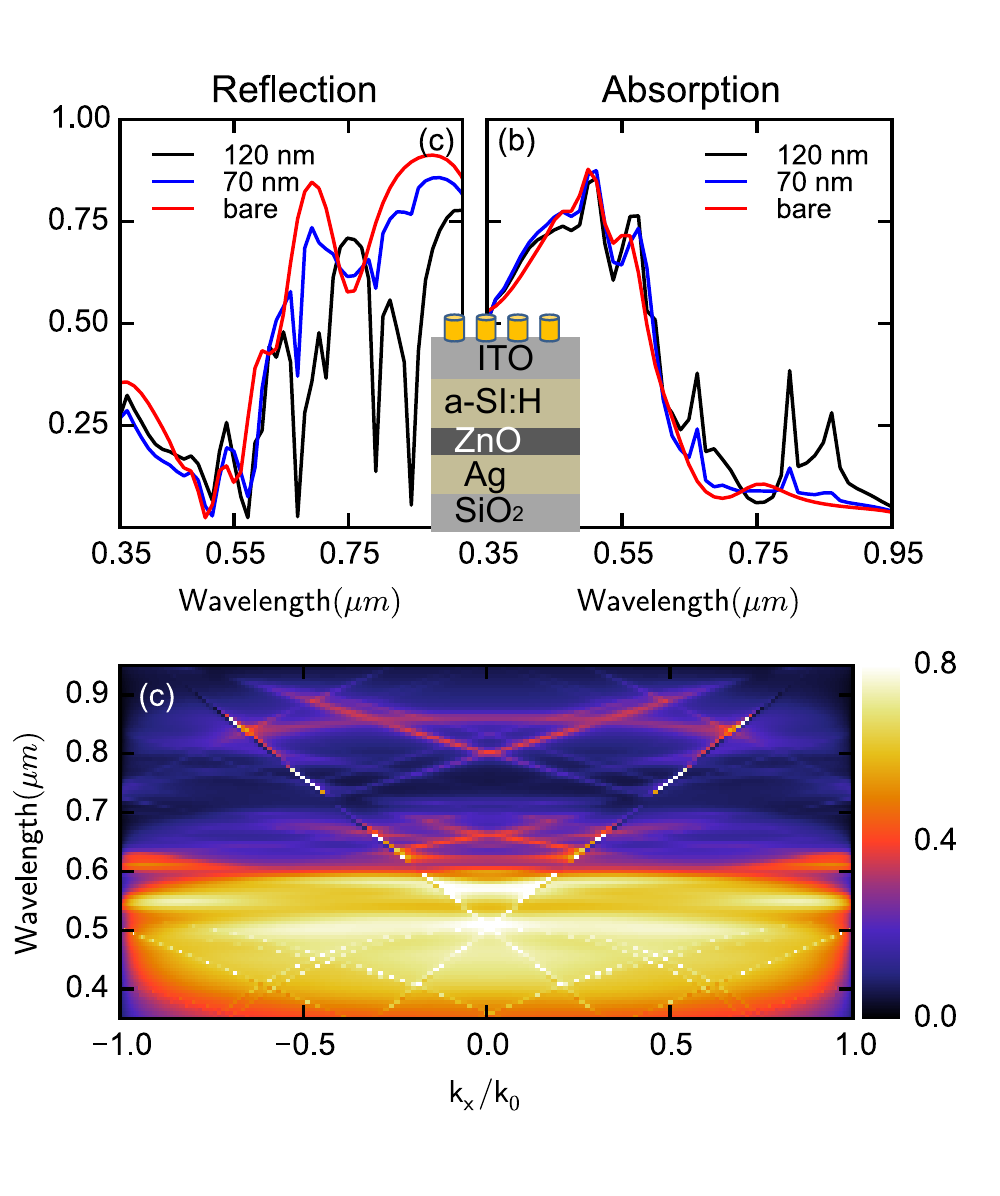}
\caption{\label{fig4} (a) Calculated reflection from the layer geometry shown in the inset, which reflects the cell geometry of a realistic photovoltaic cell based on amorphous silicon~\cite{LareAPL2012}. (b) Calculated absorption in the a-Si:H i-layer.  Red line for the bare layer geometry without plasmonic lattice, blue line  for the plasmonic lattice with particle radius of 70 nm, black line for the plasmonic lattice with particle radius of 120 nm. (c) Absorption  as function of incident angle and  photon energy. In (a-b),  light is illuminated at normal incidence. In (c), the incident light has TE polarization.}
\end{figure}

%

As second example of the utility of our method  we apply our dipole theory to  a realistic photovoltaic (PV) structure, in which the plasmonic lattice is coupled to thin-film silicon solar cells containing several layers, as studied by van  Lare \cite{LareAPL2012} et. al. In particular, the application area is to amorphous silicon (a-Si:H) photovoltaics, in which 350 nm of silicon is sandwiched between an ITO contact, on one side, and a 80 nm ZnO buffer layer to a 200 nm thick silver back contact on the other side. One challenge of such cells is that the silicon is much thinner than the absorption length, especially in the red. Consequently, these cells have up to 80\% reflectivity for wavelengths above 650 nm. The aim of plasmonic structuring is  to improve absorption in the amorphous silicon layer, despite the fact that it is far thinner than the bulk absorption length, by coupling into guided modes. This example  illustrates the potential application of our method to rapidly screen complex photonic structures in  thin-film photovoltaics. In the construction of our dyadic Green's function, we implement  the full layer-structure of  the  thin-film silicon solar cells. However,  to a good approximation, we take the 200 nm thick Ag back contact as infinite.   We use tabulated optical constants for all materials, taken from \cite{LareAPL2012,Palik}.

\Figure{fig4} (a,b) shows the calculated reflection, absorption in a-Si:H i-layer for the  bare layer geometry, and geometries that include embedded  plasmonic particle lattices of pitch 500 nm, and two different particle sizes (radius of 70 nm, and 120 nm). In \Fig{fig4} (a), it is evident that without the plasmonic lattice, the cell shows minimum reflectivity at 550 nm, and a large 80\% reflectivity for wavelengths above 650 nm. This is in good accord with the full wave simulations of van Lare, and highlights a main problem of thin film amorhous solar cell, namely that light is reflected at the back contact after a single pass, without being absorbed. The plasmonic lattices strongly reduce the cell reflection in a set of narrow spectral lines, where the introduced reflection minima are most pronounced for the larger particles, again in good accord with the work of van Lare \cite{LareAPL2012}.  \Fig{fig4} (b) shows the  absorption inside the amorphous silicon material only. The plasmonic structure introduces pronounced absorption peaks    at wavelengths near the reflection minima, that are due to diffractive coupling into waveguide modes.

To further quantify the role of the diffractive coupling between the guided modes and the free propagating modes in enhancing absorption, we study an $\omega$-k diagram of the absorption \cite{Ferry2010OE}. \Figure{fig4} (c) shows the calculated absorption in the a-Si:H i-layer for the case where a plasmonic lattice with particles of radius 120 nm is embedded in the system, as function of incident angle (with TE polarization) and photon energy.  As in \Fig{fig4} (a,b), the absorption diagram as function of frequency and wave vector shows sharp and dispersive features on top of a background that is independent of angle, that shows a sharp step in absorption at 0.6 $\mu$m wavelength. This sharp step  is due to a sharp change in  absorption coefficients of a-Si:H layer at 0.6 $\mu$m. The very sharp and dispersive features are due to diffractive resonances, i.e, diffractive coupling to the guided modes in the a-Si:H layer, as well as  lattice resonances given by the sharp bright lines. Very similar wavelength and angle-dependent features in photocurrent have been measured for actual cells by Ferry et al. \cite{Ferry2010OE},and such measurements can be an important tool to unravel the physics behind photovoltaic performance improvements. Beyond the relevance of this particular amorphous silicon cell example, the calculations evidence that  our method can function as a rapid means to screen different light trapping strategies without the need for full wave simulations.


\subsection{Excitation of dark  modes by point sources\label{subsec:dark}}

\begin{figure}
\includegraphics[width=0.7\textwidth]{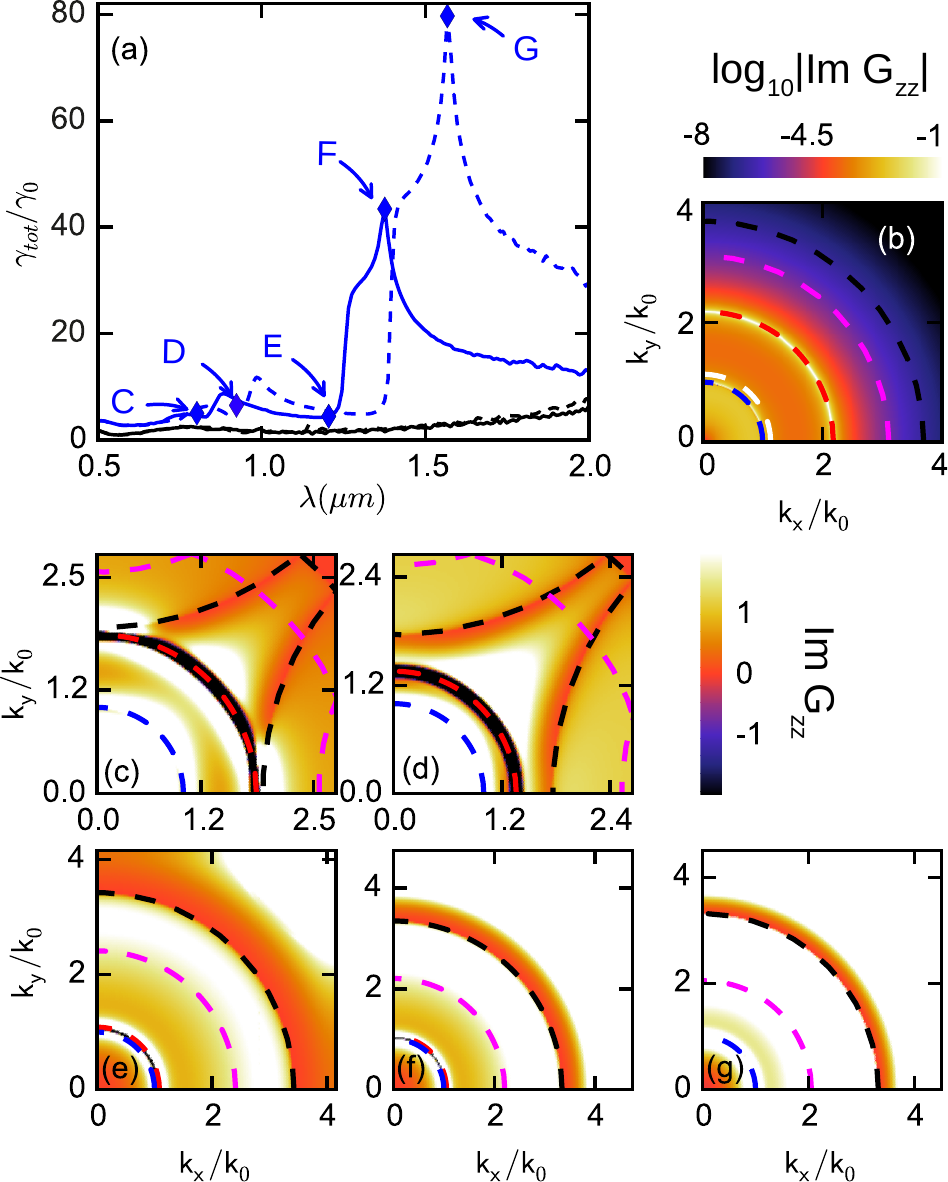}
\caption{\label{fig5} Emission rate of a dipole coupled to an aluminum sphere array embedded in a GaAs slab waveguide for an x-oriented dipole (black curves, left y-axis), and for a z-oriented dipole (blue curves, righthand y-axis). Solid (dashed) curves are for a lattice pitch of 145 (173) nm.  (b) Contour plot of the integrand of a bare slab Green function $\log_{10}|Im [ G_{zz}^{s}]|$. (c-g) Contour plot of    $Im [ G_{zz}]$, i.e., $\bar { \bm{ I } }(\bm k_{||},\bm r,\bm r_0)$ in \Eq{master_equation}  for the special points, as marked by C,D, E, F, G in (a). In (b-g), the blue (black) dotted lines indicate the folded light line in air (GaAs), the white and magenta dotted lines for TE bound modes, the red dotted line for TM bound mode.}
\end{figure}

%

For many interesting quantities, plane wave calculations, together with the reciprocity technique suffice.  These include angle-dependent extinction, absorption, reflection, refraction and diffraction and the differential radiated power for a localized source to any far field angle. However, for some highly interesting quantities, the array scanning method is indispensable.  In general,  a photonic structure may offer guided modes as efficient channels for emission,  or in case of a plasmonic structure, may cause absorption. Generally, the total decay rate enhancement is due to the sum of induced quenching, emission into guided modes, and an increased rate of emission into far field modes. Only the last term is accessible to the reciprocity technique.  The array scanning method, on the other hand, can be used to unravel quenching, and guided mode contributions.  The full Brillouin zone integral in the array scanning method provides the sum of all three contributions, i.e., the full local density of states. In addition, as it gives the field at any position due to a localized source, one can calculate absorption.  Finally,  analyzing the integrand of \Eq{master_equation} provides insight in the specific wave vectors that contribute to the local density of states, for instance pointing at guided mode contributions to the emission.

As third example to illustrate our method, we discuss a fluorescence problem tackled by the ASM method in which part of the emission enhancement of a dipole source is funneled into guided modes.  As geometry we take a thin ($d=116 nm$) layer of GaAs  in air,  and assume that a square lattice (lattice constant $p$) of aluminum particles (radius 0.25$p$) is embedded in the slab at about 2/3rds of the slab thickness (all dielectric constants taken from Palik \cite{Palik}).  We assume a source located just above the slab top surface (25 nm separation).  This geometry is not altogether realistic from an experimental stand point but has many illustrative challenges from a computational point of view.  These are (A) as opposed to the previous examples the particles are \emph{inside} a layer while the source is on top, (B)   strong coupling of the particles to the slab waveguide mode, (C) no mirror symmetry in the geometry, so that particles couple to TE and TM modes, and   (D) evanescent coupling of the source to the slab wave guide mode.   We consider this computationally tough problem in a range of near- to mid-IR wavelengths, and for two pitches.

As shown in \Fig{fig5} (a),  we study the total decay rate enhancement (in current literature often referred to as Purcell factor  ($F$)) of a dipolar emitter, coupled to the scatterer/waveguide system. The total decay rate enhancement  is given by   the total emission rate $\gamma_{tot}$ divided by the vacuum rate $\gamma_{0}$. In \Fig{fig5} (a), results are shown for  the emitter oriented in plane (along x-direction, black lines), and normal to the plane (along z-direction, blue lines).  In this diagram solid lines refer to $p$=145 nm, while dashed lines refer to $p=$173 nm. For x-dipoles, the Purcell factor  increases moderately, towards long wavelengths, which is due to increased material losses in aluminum particles at long wavelength.
In contrast, the increase of the Purcell factor for a z-dipole coupled to the same configuration of  lattice-waveguide system is  pronounced, attaining values up to 40 or 80 for different pitches, as evident by the sharp peaks denoted by F (1.37 $\mu$m) and G (1.57 $\mu$m) in \Fig{fig5} (a).  The large enhancement of $\gamma_{tot}$ results from the excitation of dark modes, i.e., guided modes,  of the plasmonic lattice for long wavelengths, rather than merely from increased material losses. This is evident from two facts: (1) the peaks F, G are not close to the wavelength of maximum material losses, (2) the enhanced emission rate, up to 80,  for z-dipoles far exceeds that for x-dipoles, which is less than 8.

To further study the dark mode excitation, we plot the integrands of the scattering part of the Green function, i.e., $\bar { \bm{ I } }(\bm k_{||},\bm r,\bm r_0)$ in \Eq{master_equation}, in one quadrant of the full Brillouin zone in \Fig{fig5} (d-k)  for special points C,D, E, F, G marked in  \Fig{fig5} (a), in all cases focusing on the $z$-oriented dipole case. For reference, one can also evaluate the integrand for a `bare'  slab, containing a lattice of particles of zero polarizability,  shown in \Fig{fig5} (b)   at  wavelength of  1 $\mu$m. In all panels, the plot range encompasses exactly the positive quadrant of the first Brillouin zone. In (b-h), blue (black) dotted lines indicate the light line in air (GaAs). Due to C$_4$ symmetry of the square lattice, the integrands in the full Brillouin zone  can be obtained by mirroring along $k_x=0$, and $k_y=0$. In order to directly compare the k-space spectrum content, we use the same colormap for all  contour plots in (c-g). The bare slab system supports three waveguides modes at short wavelength  as shown in (b), i.e., two TE modes indicated by white and magenta dashed lines, plus one TM mode indicated by a  red dashed line. At longer wavelength the fundamental TM /TE mode indicated by the  magenta/red dotted line remains, while the higher order TE mode (white dotted line) vanishes.

In the bare slab waveguide in  \Fig{fig5} (c), the contribution of the waveguide mode to the integrand of scattering part of Green function is weak, as shown in the contour plot of  $\log_{10}|Im [ G_{zz}]|$. For larger in-plane momentum, the integrand  of the scattering part of Green function approaches 0. However, with the lattice of scatterers in place, the integrands of the second term in \Eq{master_equation}  are strongly mediated by the periodic scatterers.  At short wavelengths, i.e., (c,d), there are three major features: (1) the bare slab waveguide supports well-defined waveguide modes, indicated by the dark arcs; (2) diffraction occurs, where the guided modes of the slab waveguide folds back into the Brillouin zone; (3) dark plasmonic modes  are  excited on the lattice in \Fig{fig5} (d-h) in a limited region of k-space  close to the folded guided mode and is evident as very large integrand (bright features). At longer wavelengths (e-g), the dipole lattices become denser with respect to vacuum wavelength, so that back-folding no longer leads to intersections appearing with back-folded waveguide modes. Plasmon hybridization becomes stronger, and hence lattice resonances may play a significant role in the enhancement of $F$. Indeed, the integrand in \Fig{fig5} (e-g) shows that there are considerable plasmonic contributions (broad bright features) in the Brillouin zone within  the (folded) light cone ( roughly 3.35$k_0$) given by GaAs, besides the narrow contribution from guided modes, which are only weakly confined at these long wavelengths. In \Fig{fig5} (d), a major part of the  emission   is  inside the folded light cone given by the slab medium (GaAs), and includes a considerable amount inside the vacuum light cone ($k_0$), which indicates  that a significant fraction of the total emission can be delivered into the far field. In contrast, one notices that in  the integrand in \Fig{fig5} (e-g), the major contribution to $\gamma_{tot}$ lies   outside the vacuum light cone. Hence, most of the emission is funneled into the dark modes of the lattice, and is eventually dissipated due to material losses.

The emission distribution in k-space, i.e., the fraction of  the emission fuelled into far-field, as well as into the dark modes,  for points C-G are tabulated in \Table{tab1}. As evident in \Fig{fig5} (c-e), the the major contribution to the integrand appears at parallel wave vectors beyond the light-line of the host material of the lattice. The fraction of emission into the far field, i.e.,  $|\bm k_{||}|<k_0$, reaches its maximum value ($0.35\%$) at the point marked D. Further to the red, i.e, at point E, we see that emission with large parallel k-component, i.e.,  $|\bm k_{||}|\geq n_{GaAs} k_0$,  starts to contribute due to the stronger plasmonic hybridization. As the wavelength increases further, the emission with $|\bm k_{||}|\geq n_{GaAs} k_0$ quickly dominates, which yields nearly $100\%$ emission funnelled  into the dark mode of the plasmon lattice. Due to the large $|\bm k_{||}|$, those modes are tightly confined by the plasmonic lattice, and quickly dissipate due to material losses.  It is important to realize that designing the k-space emission distribution is highly non-trivial, and the rules  behind such type of design are beyond the scope of this paper. Essentially, it is the competition among  channels, i.e.,  the emission with $|\bm k_{||}<k_0$,  $k_0<|\bm k_{||}|\geq n_{GaAs} k_0$, and $|\bm k_{||}|\geq n_{GaAs} k_0$, that ultimately determines how emission occurs in the complex photonic environments~\cite{ZhouLei2015PRL}.  Our method can aid evaluating different designs on basis of the fact that the integrand in the ASM method can be separated in different contributions. To conclude this example,  the ASM-method provides key insights in the emission distribution in k-space, which could  be utilized in many problems that require engineering the emission distribution in k-space as well as the separation of emission into far field, guided modes, and absorption.

\begin{table}[t]
\centering
\begin{tabular}{cccccc}
  \hline
 \hline
     & C & D & E &  F & G \\
  $\gamma_{tot}$ &   5.04 &    6.53 &   4.47  &  43.44  &  79.69  \\
 Fraction ($|\bm k_{||}| < k_0$ ) &  0.16 &   0.35   & 0.05  &  0.005   & 0.003 \\
  Fraction  ($|\bm k_{||}|\geq n_{GaAs} k_0$)  & 0 &   0 &   0.14  &  0.92 &   0.95 \\
  \hline
    \hline
\end{tabular}
\caption{Fraction of emission of z-dipole coupled to waveguide-lattice system in k-space.   C, D, E, F, G denote the special points  marked in \Fig{fig5} (a).}
\label{tab1}
\end{table}

\section{Conclusion}
In closing, we presented a general point dipole theory for periodic metasurfaces that are embedded in arbitrary stratified layer systems. With the assistance of Ewald's summation technique and dyadic Green's function of stratified layers, we study magnetoelectric  scattering lattices coupled to  layered photonic  structures. One can with this method rapidly understand the physics of complicated structures by approximating  meta-atoms as point dipoles with magnetoelectric responses, while accounting for all particle-particle interactions self-consistently via freely propagating photons as well as bound modes provided by the stratified layers. We further extended the plane wave excitation case to excitation by a single, localized,  dipolar source, through the array scanning method.   Thereby, all the relevant optical properties, such as LDOS and far field emission patterns, of the coupled waveguide lattice, can be extracted straightforwardly.

As applications of the proposed method, we demonstated that it applies readily to solid-state lighting scenarios where point emitters are coupled to periodic plasmonic lattices, in turn embedded in high index layered systems. Also, we applied our method to a plasmon enhanced thin-film solar cell structure containing a plasmonic nanoantenna array. We find that our method can efficiently handle such  structures, capturing salient features due to diffractive coupling in  terms of extinction, absorption and reflection. Due to high efficiency and flexibility of the dipole theory, it can be used for both diffractive, dilute lattices, and for  dense lattices with complex clusters of inclusions that form a metamaterial or metasurface. We envision that our method is useful for studying metasurfaces, as well as related applications such as light-emitting diodes and solar cells.

\section{Acknowledgement}
Y.T. Chen would like to thank Prof. S. A. Tretyakov for fruitful discussions. Y.T. Chen acknowledges financial support from the National Natural Science Foundation of China (Grant No. 61405067), and Foundation for Innovative Research Groups of the Natural Science Foundation of Hubei Province (Grant No. 2014CFA004). This work is part of the research program of the ``Foundation for Fundamental Research on Matter (FOM)'', which is financially supported by the ``The Netherlands Organization for Scientific Research (NWO)''. A.F.K. gratefully acknowledges an NWO-Vidi and Vici grant for financial support.
\appendix

\section{Construction of free space  dyadic Green's function based on plane wave expansion}
For a given wave vector $\bm k= [\bm k_{||}, k_z]$, the TE and TM polarization modes associated with a plane wave are given by  $\bm E_{TE}(\bm r)=\bm M(\bm k_{||}, \pm_{1}, \pm_{2}) e^{i\bm k_{||} \cdot \bm{r}_{||}}$,  and $\bm E_{TM}(\bm r)=\bm N(\bm k_{||}, \pm_{1}, \pm_{2})e^{i\bm k_{||} \cdot \bm{r}_{||}}$  respectively, where $\bm M$ and $\bm N$ are defined as follows,
\begin{subequations} \label{TETM0000}
\begin{align}
\bm M(\bm k_{||,\pm_1,\pm_2}) & =e^{i\pm_{2} k_z z} \left(\hat{x}\frac{- k_y}{ k_{||}} +\hat{y}\frac{ k_x}{ k_{||}}\right), \\
\bm N(\bm k_{||,\pm_1,\pm_2}) & =e^{i \pm_{2} k_z z} \left[\hat{z}\frac{ i k_{||} }{k_0} - \hat{x}\frac{i  (\pm_{1}) k_z k_x}{ k_{||} k_0} -\hat{y}  \frac{i (\pm_{1} )k_z k_y}{ k_{||} k_0}  \right],
\end{align}
\end{subequations}
where $\bm r =[\bm r_{||}, z]$,  $k_{||}=|\bm k_{||}|$, $k_0=\sqrt{k_{||}^2+k_z^2}$.  The first sign factor describes the propagation direction along z-axis, while the second sign factor denotes the sign of $k_z$ in the construction of the basis function. Throughout the paper, we will use $\bm M(\bm r, \pm, \pm)$ and $\bm N(\bm r, \pm, \pm)$ for short. In the above formulation, the prefactors of TE and TM waves are arranged to fullfil the relation of
\beq\label{TETMrelation}
\begin{array}{c}
  \nabla  \times [  e^{i\bm k_{||} \cdot \bm{r}_{||}} M(\bm k_{||}, \pm, \pm) ]=  e^{i\bm k_{||} \cdot \bm{r}_{||}} k_0 N(\bm k_{||}, \pm, \pm), \\
    \nabla  \times [ e^{i\bm k_{||} \cdot \bm{r}_{||}} N(\bm k_{||}, \pm, \pm) ]= e^{i\bm k_{||} \cdot \bm{r}_{||}}  k_0  M(\bm k_{||}, \pm, \pm).
\end{array}
\eeq
In free space, the dyadic Green function of the electric field can be expanded using the plane waves as follows (upper sign applies for $z>=z'$),
\begin{eqnarray} \label{free_green}
\begin{aligned}
\bar { \bm{ G } }_0^{ee}(\bm r, \bm r') &= \frac{i}{8 \pi^2 }\int dk_xdk_y \frac{e^{i\theta} }{k_z} [\bm M_{+} \otimes \bm M'_{-} -\bm N_{+} \otimes \bm N'_{-}]  \\
\bar { \bm{ G} }_0^{he}(\bm r, \bm r') & = \frac{ik_0}{8 \pi^2 }\int dk_xdk_y   \frac{e^{i\theta} }{k_z}[\bm N_{+} \otimes \bm M'_{-} - \bm M_{+} \otimes \bm N'_{-}],\\
\bar { \bm{ G } }_0^{hh}(\bm r, \bm r') &= \frac{i}{8\pi^2 } \int dk_xdk_y \frac{e^{i\theta} }{k_z} [-\bm N_{+} \otimes \bm N'_{-}+\bm M_{+} \otimes \bm M'_{-} ] ,\\
\bar { \bm{ G} }_0^{he}(\bm r, \bm r') & = \frac{ik_0 }{8 \pi^2 } \int dk_xdk_y  \frac{e^{i\theta} }{k_z} [- \bm M_{+} \otimes \bm N'_{-}+\bm N_{+} \otimes \bm M'_{-}].
\end{aligned}
\end{eqnarray}
where $\theta=\bm k_{||} \cdot (\bm r_{||}-\bm r_{||}^{'})$, $\bm M_{(\pm)}/\bm N_{(\pm)}=\bm M/\bm N(\bm k_{||}, \pm,\pm)$, $\bm M'_{(\mp)}/\bm N'_{(\mp)}=\bm M/\bm N(\bm k_{||}, \pm,\mp)$.

\section{Dyadic Green's function for an interface}
Dyadic Green function for an interface or  stratified structure   can be expanded using the plane waves, i.e., $\bar{\bm{G} }_{if}(\bm r, \bm r')=\bar{\bm{G} }_{0}(\bm r, \bm r')+\bar{\bm{G} }_{s}^{if}(\bm r, \bm r') $, where $\bar{\bm{G} }_{0}(\bm r, \bm r')$ is the free space terms given by \Eq{free_green}.

%
 \begin{eqnarray} \label{free_green}
\begin{aligned}
  \bar { \bm{ G } }_{s,ee}^{if}  & = \frac{i}{8 \pi^2 }\int d\bm{k}_{||}   \frac{e^{i\theta} }{ k_z} \{\gamma_{s}  \bm M_{+}\otimes  \bm M'_{+} -\gamma_{p}  \bm N_{+} \otimes \bm N'_{+} \} \\
   \bar { \bm{ G} }_{s,he}^{if}  & = \frac{ik_0 }{8 \pi^2}\int d\bm{k}_{||}  \frac{e^{i\theta} }{ k_z}  \{\gamma_{s}   \bm N_{+}\otimes \bm M'_{+}  - \gamma_{p}  \bm M_{+} \otimes \bm N'_{+}\},\\
  \bar { \bm{ G } }_{s,hh}^{if}  &= \frac{i }{8 \pi^2 } \int d\bm{k}_{||}  \frac{e^{i\theta} }{ k_z} \{\gamma_{p}  \bm M_{+} \otimes \bm M'_{+}-\gamma_{s}  \bm N_{+}\otimes \bm N'_{+}  \},\\
  \bar { \bm{ G} }_{s,eh}^{if}  & = \frac{ik_0}{8 \pi^2 }\int d\bm{k}_{||}  \frac{e^{i\theta} }{ k_z}  \{ \gamma_{p}  \bm N_{+} \otimes \bm M'_{+}-\gamma_{s}   \bm M_{+}\otimes \bm N'_{+}\}.
\end{aligned}
\end{eqnarray}
\subsection{Dyadic Green function for an slab structure}
Dyadic Green function for an slab structure can be expanded using the plane waves as follows \cite{KlimovPRA2005, HartmanJCP,Cheng1986}, see the geometry in \Fig{fig1} (a), in which both the emitter and the detection point are located in the layer 2,
\begin{eqnarray} \label{slab_green}
\begin{aligned}
&  \bar { \bm{ G } }^{ee}_{_{2,2}}(\bm r, \bm r')   = \int d\bm {k}_{||} \frac{ie^{i\theta}}{8 \pi^2 k_{z2}} \{  \\&  \frac{1}{\Delta_s}[\bm M_{(+)} + \gamma_{s,21}(k_{||})\bm M_{(-)}\eta] \otimes [ \bm M^{'}_{(-)} + \gamma_{s,23}(k_{||}) \bm M^{'}_{(+)}] \\
  &-\frac{1}{\Delta_p}[ \bm N_{(+)} +  \gamma_{p,21}(k_{||})\bm N_{(-)} \eta] \otimes [ \bm N^{'}_{(-)} +\gamma_{p,23}(k_{||}) \bm N^{'}_{(+)}]  \} ,\\
&  \bar { \bm{ G } }^{he}_{_{2,2}}(\bm r, \bm r')   = \int d\bm{k}_{||} \frac{ie^{i\theta}k_0}{8 \pi^2 k_{z2}} \{  \\& \frac{1}{\Delta_s}[ \bm N_{(+)} + \gamma_{s,21}(k_{||})   \bm N_{(-)}\eta] \otimes [ \bm M^{'}_{(-)} + \gamma_{s,23}(k_{||}) \bm M^{'}_{(+)}] \\
  &-\frac{1}{\Delta_p}[   \bm M_{(+)} +  \gamma_{p,21}(k_{||})   \bm M_{(-)} \eta] \otimes [ \bm N^{'}_{(-)} +\gamma_{p,23}(k_{||}) \bm N^{'}_{(+)}]  \}, \\
 &  \bar { \bm{ G } }^{hh}_{_{2,2}}(\bm r, \bm r')  = \int d\bm{k}_{||} \frac{ie^{i\theta}}{8 \pi^2 k_{z2}} \{    \\& \frac{1}{\Delta_s}[\bm N_{(+)} + \gamma_{s,21}(k_{||})\bm N_{(-)}\eta] \otimes [ \bm N^{'}_{(-)} + \gamma_{s,23}(k_{||}) \bm N^{'}_{(+)}] \\
  &-\frac{1}{\Delta_p}[ \bm M_{(+)} +  \gamma_{p,21}(k_{||})\bm M_{(-)} \eta] \otimes [ \bm M^{'}_{(+)} +\gamma_{p,23}(k_{||}) \bm M^{'}_{(-)}]  \}, \\
& \bar { \bm{ G } }^{eh}_{_{2,2}}(\bm r, \bm r')  = \int d\bm{k}_{||} \frac{ie^{i\theta}k_0}{8 \pi^2 k_{z2}} \{   \\& \frac{1}{\Delta_s}[\bm M_{(+)} + \gamma_{s,21}(k_{||})  \bm M_{(-)}\eta] \otimes [ \bm N^{'}_{(-)} + \gamma_{s,23}(k_{||}) \bm N^{'}_{(+)}] \\
  &-\frac{1}{\Delta_p}[   \bm N_{(+)} +  \gamma_{p,21}(k_{||})   \bm N_{(-)} \eta] \otimes [ \bm M^{'}_{(-)} +\gamma_{p,23}(k_{||}) \bm M^{'}_{(+)}]  \} .
\end{aligned}
\end{eqnarray}

where  $\Delta_{s/p}=1- \gamma_{s/p,21}(k_{||}) \gamma_{s/p,23}(k_{||})e^{i2k_zd}$, and  $\gamma_{s,ij}=\frac{k_{zi}-k_{zj}}{k_{zi}+k_{zj}}$, $\gamma_{p,ij}=\frac{\epsilon_j k_{zi}-\epsilon_i k_{zj}}{k_{zi}\epsilon_j+k_{zj}\epsilon_i}$, $\eta= e^{i2k_zd}$.

\end{document}